\documentclass[twocolumn,preprintnumbers]{revtex4}

\def\be{\begin{equation}}
\def\ee{\end{equation}}
\def\beq{\begin{eqnarray}}
\def\eeq{\end{eqnarray}}
\def\n{\nonumber}
\def\bay{\begin{array}}
\def\eay{\end{array}}

\begin{document}

\preprint{CIRI/02-sw04}
\title{Inhomogeneous Big Bang Cosmology}

\author{Sanjay M. Wagh}
\affiliation{Central India Research Institute, Post Box 606,
Laxminagar, Nagpur 440 022, India\\
E-mail:cirinag@nagpur.dot.net.in }

\date{November 3, 2002}
\begin{abstract}
In this letter, we outline an inhomogeneous model of the Big Bang
cosmology. For the inhomogeneous spacetime used here, the universe
originates in the infinite past as the one dominated by vacuum
energy and ends in the infinite future as the one consisting of
``hot and relativistic" matter. The spatial distribution of matter
in the considered inhomogeneous spacetime is {\em arbitrary}.
Hence, observed structures can arise in this cosmology from
suitable ``initial" density contrast. Different problems of the
standard model of Big Bang cosmology are also resolved in the
present inhomogeneous model. This inhomogeneous model of the Big
Bang Cosmology predicts ``hot death" for the universe.
\\
\centerline{Submitted to: ------------------------------ }
\end{abstract}
\maketitle

\newpage
\section{Introduction} In the 3+1 formulation \cite{mtw} of
General Relativity,  {\em source}, matter or energy, data can be
specified on some suitable ``initial" spacelike hyper-surface and,
that data can be evolved using the Einstein field equations.
Different four-dimensional spacetime geometries are obtainable for
different initial source data.

When matter or energy sources are present over all of the initial
hyper-surface, we obtain a ``cosmological"  situation or a
``cosmological" spacetime. A famous is the case of maximally
symmetric, homogeneous and isotropic, Friedmann-Lema\'\i
tre-Robertson-Walker (FLRW) spacetime of the Big Bang Cosmology
(BBC) \cite{mtw, sussp, stdbooks}.

Now, to fix ideas, recall that the Newtonian Law of Gravitation is
a statement of the force of attraction between two mass points in
space. The presence of other mass-points does not alter the
statement of this Law of Gravitation. Moreover, re-distribution of
mass-points does not affect the statement of this Law of
Gravitation.

General Relativity prescribes a spacetime geometry for a Law of
Gravitation. In analogy with the Newtonian case, we therefore seek
a spacetime which ``does not change" its geometrical characters
when the source distribution in it is altered in some way.
Moreover, the requirement that the addition of ``more" sources
does not change the geometrical characters of this spacetime means
that it is also a ``cosmological" spacetime.

But, we note that the FLRW spacetime does {\em change\/}  its
geometrical characters when matter is differently distributed,
say, in some inhomogeneous way, in this spacetime. Therefore, it
does not satisfy our above criteria.

Hence, we consider here a Petrov-type D, cosmological spacetime,
of (\ref{genhsp}), which ``does not change" its geometrical
characters under redistribution of matter in it and explore its
cosmology, broadly the nature of the dynamics of the
``cosmological spacetime" under consideration.

We recall that the, homogeneous and isotropic, FLRW metric
displays \cite{mtw, sussp, stdbooks} different dynamical behavior
for differing source constituents in it. Similarly, the
inhomogeneous spacetime of (\ref{genhsp}) displays different
dynamical behavior for differing source contents in it. In this
letter, we focus on the inhomogeneous realization of the big bang
cosmology using the spacetime of (\ref{genhsp}).

\section{The Cosmological Spacetime} \label{spacetime} Consider the
spacetime metric: \beq ds^2 &=& -X^2Y^2Z^2dt^2 \n \\ &\phantom{=}&
\quad +\gamma^2{X'}^2Y^2Z^2B^2 dx^2 \n \\
&\phantom{=}& \qquad +\;\gamma^2X^2\bar{Y}^2 Z^2C^2dy^2 \n \\
&\phantom{=}& \quad\qquad +\;\gamma^2X^2Y^2\tilde{Z}^2D^2\,dx^2
\label{genhsp} \eeq where $X\equiv X(x)$, $Y\equiv Y(y)$, $Z\equiv
Z(z)$, $B\equiv B(t)$, $C\equiv C(t)$, $D\equiv D(t)$ and $\gamma$
is a constant. Also, $X'=dX/dx$, $\bar{Y}=dY/dy$ and
$\tilde{Z}=dZ/dz$.

\subsection*{Singularities and Degeneracies of (\ref{genhsp})}
There are, in general, two types of singularities of the metric
(\ref{genhsp}). The first type is when any of the temporal
functions $B$, $C$, $D$ is vanishing and the second type is when
any one of the spatial functions $X$, $Y$, $Z$ is vanishing.

Singularity of the first type, {\it ie}, vanishing of temporal
function, corresponds to a singular hyper-surface of
(\ref{genhsp}). On the other hand, singularities of the second
type constitute a part of the initial data, singular spatial data,
for (\ref{genhsp}).

The locations for which the spatial derivatives vanish are,
however, coordinate singularities. The curvature invariants of
(\ref{genhsp}) do not blow up at such locations. Therefore, before
such coordinate singularities are reached, we may transform
coordinates to other suitable ones.

There are also obvious degenerate metric situations when any of
the spatial functions is infinite for some range of the
coordinates.

{\em In what follows, we shall assume that there are no singular
spatial initial data and that there are no spatially degenerate
situations for the cosmological spacetime of the metric
(\ref{genhsp}).}

\subsection*{Other features of the metric (\ref{genhsp})}
The metric (\ref{genhsp}) admits three hyper-surface orthogonal
spatial homothetic Killing vectors (HKVs) \beq {\bf {\cal X}}
&=& (0, \frac{X}{\gamma X'}, 0, 0) \label{genhkv1} \\
{\bf {\cal Y}} &=& (0, 0, -\frac{Y}{\gamma \bar{Y}}, 0) \label{genhkv2} \\
{\bf {\cal Z}} &=& (0, 0, 0, -\frac{Z}{\gamma\tilde{Z}})
\label{genhkv3} \eeq

Now, the existence of three HKVs is equivalent to two Killing
vectors (KVs) and one HKV. Then, the metric (\ref{genhsp}) can be
expressed in a form that displays the existence of KVs explicitly.
However, we will use the form (\ref{genhsp}) in this letter.

The spacetime of (\ref{genhsp}) is required, by definition, to be
locally flat at all of its points. In general, this will require
some conditions on the metric functions $X$, $Y$, $Z$.

Also, the non-vanishing components of the Weyl tensor for
(\ref{genhsp}) are: \beq C_{txtx} &=&
\frac{B^2\gamma^2X'^2Y^2Z^2}{6}\,F(t) \label{weyl1}
\\
C_{tyty} &=& \frac{C^2\gamma^2X^2\bar{Y}^2Z^2}{6}\,G(t) \label{weyl2} \\
C_{tztz} &=& \frac{D^2\gamma^2X^2Y^2\tilde{Z}^2}{6}\,H(t)
\label{weyl3} \\ C_{xyxy} &=&
-\,\frac{B^2C^2\gamma^4X'^2\bar{Y}^2Z^2}{6}\,H(t)\label{weyl4} \\
C_{xzxz} &=& -\,\frac{B^2D^2\gamma^4X'^2Y^2\tilde{Z}^2}{6}\,G(t)
\label{weyl5} \\ C_{yzyz} &=& -\,
\frac{C^2D^2\gamma^4X^2\bar{Y}^2\tilde{Z}^2}{6}\,F(t)
\label{weyl6} \eeq where \beq F(t) &=& \frac{\ddot{C}}{C}
+\frac{\ddot{D}}{D} -2\frac{\ddot{B}}{B} \n \\ &\phantom{=}& -
2\frac{\dot{C}\dot{D}}{CD}
+\frac{\dot{B}\dot{C}}{BC} + \frac{\dot{B}\dot{D}}{BD}  \\
G(t) &=& \frac{\ddot{D}}{D} +\frac{\ddot{B}}{B}
-2\frac{\ddot{C}}{C}  \n \\ &\phantom{=}&  +
\frac{\dot{C}\dot{D}}{CD}
+\frac{\dot{B}\dot{C}}{BC} -2 \frac{\dot{B}\dot{D}}{BD} \\
H(t)&=& \frac{\ddot{B}}{B} +\frac{\ddot{C}}{C}
-2\frac{\ddot{D}}{D} \n \\ &\phantom{=}& +
\frac{\dot{C}\dot{D}}{CD} +\frac{\dot{B}\dot{D}}{BD} -2
\frac{\dot{B}\dot{C}}{BC} \eeq

Now, for (\ref{genhsp}), it can be verified that the non-vanishing
Newman-Penrose (NP) complex scalars \cite{mtbh} are $\Psi_0 =
\Psi_4$ and $\Psi_2$.

Note that $\Psi_4 \neq 0$. Now, consider a NP-tetrad rotation of
Class II, with complex parameter $b$, that leaves the NP-vector
${\bf n}$ unchanged, and demand that the new value,
$\Psi_0^{(1)}$, of $\Psi_0$ is zero. Then, we have: \be\Psi_4\,b^4
+ 6 \Psi_2\,b^2 + \Psi_4 = 0 \ee This equation has {\em two\/}
distinct double-roots for the complex parameter $b$. Thus, the
spacetime of (\ref{genhsp}) is of Petrov-type D.

It is well-known that Penrose \cite{penroseweyl} is led to the
Weyl hypothesis on the basis of thermodynamical considerations, in
particular, those related to the thermodynamic arrow of time. On
the basis of these considerations, we may consider the Weyl tensor
to be ``measure" of the {\em entropy\/} in the spacetime at any
given epoch.

Now, note that the Weyl tensor, (\ref{weyl1}) - (\ref{weyl6}),
blows up at the singular hyper-surface of (\ref{genhsp}) but is
``finite" at other hyper-surfaces for non-singular and
non-degenerate spatial data. Then, the {\em entropy\/} at the
singular hyper-surface of (\ref{genhsp}) is ``infinite" while that
at other hyper-surfaces is ``finite". Thus, the spacetime of
(\ref{genhsp}) has the ``right" kind of thermodynamic arrow of
time in it.

The four-velocity of the co-moving observer is $U^a =
{\delta^a}_t/XYZ$. The expansion for (\ref{genhsp}) is: $\Theta
\propto \dot{B}/B + \dot{C}/C + \dot{D}/D$.

Now, the co-moving 4-velocity of matter is \be
u^a\,=\,\frac{1}{XYZ\sqrt{\Delta}}\left(1, V^x, V^y, V^z\right)
\ee where $V^x = dx/dt$, $V^y = dy/dt$, $V^z = dz/dt$ and \beq
\Delta &=& 1 - \gamma^2\left[\frac{X'^2B^2{V^x}^2}{X^2}
 \right. \n \\ &\phantom{=}& \qquad
\left. +\frac{\bar{Y}^2C^2{V^y}^2}{Y^2} +
\frac{\tilde{Z}^2D^2{V^z}^2}{Z^2} \right] \label{delta} \eeq

Then, it can be inferred \cite{physical} that the co-moving
velocity of matter is the speed of light at the singular
hyper-surface of (\ref{genhsp}). At the singular hyper-surface of
(\ref{genhsp}), $\Delta =1$. Then, a co-moving observer also moves
with the speed of light at the singular hyper-surface.

After all, matter everywhere becomes relativistic as different
mass condensates grow (due to accretion) to influence the entire
spacetime to become relativistic. This is happening asymptotically
for infinite co-moving time.

Consequently, the spacetime of (\ref{genhsp}) ends in the infinite
future as a soup of high energy plasma and radiation. This is
evidently consistent with the temporal behavior of the Weyl tensor
as a measure of the entropy.

Now, if $d\tau_{\scriptscriptstyle CM}$ is a small time duration
for a co-moving observer and if $d\tau_{\scriptscriptstyle RF}$ is
the corresponding time duration for the observer in the rest frame
of matter, then we have \be d\tau_{\scriptscriptstyle
CM}\;=\;\frac{d\tau_{\scriptscriptstyle RF}}{\sqrt{\Delta}}
\label{rshift1} \ee From (\ref{rshift1}), we also get the
red-shift formula \be \nu_{\scriptscriptstyle CM} =
\nu{\scriptscriptstyle RF}\,\sqrt{\Delta}\qquad z =
\frac{\nu{\scriptscriptstyle RF}}{\nu{\scriptscriptstyle
CM}}=\frac{1}{\sqrt{\Delta}}\label{rshift2} \ee in the spacetime
of (\ref{genhsp}) where $\nu_{\scriptscriptstyle CM}$ is the
frequency of a photon in the co-moving frame,
$\nu_{\scriptscriptstyle RF}$ is the frequency in the rest frame
and $z$ is the {\em total\/} red-shift of a photon.

Then, the red-shift, (\ref{rshift2}), depends on spatial functions
$X$, $Y$, $Z$ which determine the spatial density of matter and,
of course, on the temporal functions in (\ref{genhsp}). The
red-shift is then indicative of the contribution due to expansion
of the spacetime of (\ref{genhsp}) and the state of collapse of
matter, both.

In the past, in a different context but, it had been aptly
emphasized \cite{arp} that the red-shift of an object should have
an ``intrinsic" component, over and above that due to expansion.

Now, the Einstein tensor for (\ref{genhsp}) has the components
\beq G_{tt}&=& -\frac{1}{\gamma^2B^2}-\frac{1}{\gamma^2C^2}
-\frac{1}{\gamma^2D^2} \n \\ &\phantom{=}& \qquad\quad +
\frac{\dot{B}\dot{C}}{BC} +
\frac{\dot{B}\dot{D}}{BD} +\frac{\dot{C}\dot{D}}{CD} \label{gtt}\\
G_{xx}&=&\frac{\gamma^2B^2X'^2}{X^2}
\left[-\frac{\ddot{C}}{C}-\frac{\ddot{D}}{D} -
\frac{\dot{C}\dot{D}}{CD} \right. \n \\ &\phantom{=}& \qquad
\left. +\frac{3}{\gamma^2B^2}+ \frac{1}{\gamma^2C^2} +
\frac{1}{\gamma^2D^2}\right] \label{gxx}
\\G_{yy}&=& \frac{\gamma^2C^2\bar{Y}^2}{Y^2}
\left[-\frac{\ddot{B}}{B}-\frac{\ddot{D}}{D} -
\frac{\dot{B}\dot{D}}{BD}  \right. \n \\ &\phantom{=}& \qquad
\left. +\frac{3}{\gamma^2C^2}+ \frac{1}{\gamma^2B^2} +
\frac{1}{\gamma^2D^2}\right] \label{gyy}
\\G_{zz}&=& \frac{\gamma^2D^2\tilde{Z}^2}{Z^2}
\left[-\frac{\ddot{B}}{B}-\frac{\ddot{C}}{C} -
\frac{\dot{B}\dot{C}}{BC} \right. \n \\ &\phantom{=}& \qquad
\left. +\frac{3}{\gamma^2D^2}+ \frac{1}{\gamma^2B^2} +
\frac{1}{\gamma^2C^2}\right] \label{gzz} \\
G_{tx}&=&2\frac{\dot{B}X'}{BX}, \qquad\quad
G_{ty}=2\frac{\dot{C}\bar{Y}}{CY}\n \\
G_{tz}&=& 2\frac{\dot{D}\tilde{Z}}{DZ} \label{fluxes}\\
G_{xy}&=& 2\frac{X'\bar{Y}}{XY}, \qquad\quad
G_{xz}=2\frac{X'\tilde{Z}}{XZ}, \n \\
G_{yz}&=& 2\frac{\bar{Y}\tilde{Z}}{YZ} \label{stresses} \eeq where
an overhead dot is used to denote a time-derivative.

Now, let the energy-momentum tensor be: \beq T_{ab} &=& (\rho + p+
{\cal B})U_aU_b + (p+{\cal B})\, g_{ab} \n \\ &\phantom{=}& \quad
+ q_aU_b + q_bU_a + \Pi_{ab} \eeq where $\rho$ is the energy
density, $p$ is the isotropic pressure, $q_a = (0, q_x, q_y, q_z)$
is the energy-flux four-vector, ${\cal B}$ is the bulk-viscous
pressure, $\Pi_{ab}$ is the anisotropic stress tensor etc. Note
also that $q_aU^a = 0$ and that $\Pi_{ab}U^b = {\Pi^a}_a = 0$.
Thus, $\Pi_{ab}$, is symmetric, spatial and trace-free.

Now, from the Einstein field equations, with $8\pi G/c^4=1$, we
get $\rho = G_{tt}/X^2Y^2Z^2$. The isotropic pressure is
obtainable from (\ref{gxx}), (\ref{gyy}) and (\ref{gzz}). The
energy-fluxes $q_x$, $q_y$ and $q_z$ are related to corresponding
components of the Einstein tensor in (\ref{fluxes}) and the
anisotropic stresses are related to corresponding components of
the Einstein tensor in (\ref{stresses}).

All the spatial functions $X$, $Y$, $Z$ are {\em not\/} determined
by the field equations and, hence, are {\em arbitrary}. The
``initial temporal data" is given by chosen ``initial" values of
$\dot{B}$, $\dot{C}$, $\dot{D}$ and $\ddot{B}$, $\ddot{C}$,
$\ddot{D}$. Anisotropic stresses get determined from spatial
functions $X$, $Y$ and $Z$ as initial data.

The $x$, $y$, $z$ co-moving velocities are respectively $\propto
\dot{B},\,\dot{C},\,\dot{D}$ for (\ref{genhsp}). Therefore, the
temporal evolution in (\ref{genhsp}) leads only to a temporal
singularity in the future from ``initial" velocities and
``initial" accelerations for the non-singular and non-degenerate
spatial data.

Now, non-gravitational processes primarily determine the relation
of density and pressure of matter - the equation of state. Source
properties, as are applicable at any given stage of evolution of
source matter or energy, such as an equation of state, the
radiative characteristics etc.\ determine the temporal functions
of (\ref{genhsp}).

But, to be able to explicitly solve the corresponding ordinary
differential equations, we require {\em physical\/} information
about matter or energy. To provide for the required information of
``physical" nature is a non-trivial task in general relativity
just as it is for the Newtonian gravity \cite{stars}. The details
of these considerations are, of course, beyond the scope of the
present letter.

However, the spatial source data is, clearly, arbitrary for
(\ref{genhsp}) and the temporal evolution of non-singular and
non-degenerate data is governed by only the temporal functions in
(\ref{genhsp}).

Therefore, changing matter distribution in the spacetime does not
change the geometrical properties of (\ref{genhsp}) for
non-singular and non-degenerate data. Thus, (\ref{genhsp}) is the
metric of the spacetime that we have been looking for.

{\em Importantly, we note that there is {\em no\/} initial
spacetime singularity for (\ref{genhsp}). That is to say, there is
{\em no\/} Big Bang singularity for (\ref{genhsp})! However, the
``cosmic fireball" of the ``Big Bang" conception can occur for
(\ref{genhsp}). (See below.) Thus, the spacetime of (\ref{genhsp})
does away with the Big Bang singularity in an entirely {\em
classical\/} manner!}

As a separate remark, we note that, following the works of Ellis
and Sciama \cite{ellissciama}, there is an interpretation
\cite{tod} of Mach's principle, namely that there should be no
source-free contributions to the metric or that there should be no
source-free Weyl tensor for a Machian spacetime.

We, therefore, note that the ``absence of source" is a degenerate
case for the metric (\ref{genhsp}) and, hence, it has no
source-free contributions. Further, the spacetime of
(\ref{genhsp}) does not possess the source-free Weyl tensor. The
spacetime of (\ref{genhsp}) is, then, also a Machian spacetime in
this sense.

Now, we have sufficient characteristics of the spacetime of
(\ref{genhsp}) to discuss its cosmological implications to which
we now turn to.

\section{Cosmological implications}
The spacetime of (\ref{genhsp}) evolves based on the properties of
its source. Energy density, $\rho$, of its source has, in general,
contributions from non-relativistic matter (NRM), relativistic
matter (RM), radiation (RAD), vacuum energy (VAC), {\it ie}, \be
\rho \equiv \rho_{{\rm VAC}}+\rho_{{\rm NRM}}+\rho_{{\rm
RM}}+\rho_{{\rm RAD}}\ee etc. Then, the temporal functions of
(\ref{genhsp}) can be determined, approximately, depending on
which of these components is dominating. Detailed cosmology of
(\ref{genhsp}) is then determined in a manner similar to the
standard model \cite{mtw, sussp, stdbooks}.

Since the spatial distribution of source data is arbitrary for
(\ref{genhsp}), presently observed structures in the universe can,
clearly, evolve from some suitable ``initial" density
distribution. The entropy, of course, increases with time for
(\ref{genhsp}).

Thus, the cosmological spacetime of (\ref{genhsp}) has no
``structure formation" and no ``entropy" problems. Moreover, the
``horizon" problem is resolved by the arbitrariness of ``initial"
data for (\ref{genhsp}). ``Similar" physical characteristics
obtain for ``similar" mass condensates in it. Large-scale
homogeneity is attributable to ``initial" data for (\ref{genhsp}).

Now, for $\Theta > 0$, matter condensates in (\ref{genhsp}) are
never contracting in {\em all\/} directions. That is, there must
be, at least, one direction in which matter must be expanding
sufficiently rapidly for positive expansion, {\it ie}, $\Theta >
0$.

Without details or any justification, we only note here that
matter and energy out-flows exist in jets of radio galaxies and
quasars. We will, therefore, assume that the cosmological
spacetime of (\ref{genhsp}) is an expanding one, {\it ie}, $\Theta
> 0$ for it.

To be able to drive the expansion of the universe, we then need an
appropriate {\em source\/} which achieves this. The decaying
vacuum energy density is then an obvious candidate.

Note, however, that the field equations for (\ref{genhsp}) also
contain initial stresses. Hence, the vacuum field energy will have
to be satisfying appropriate conditions. Further, the vacuum
energy decays through particle production. For (\ref{genhsp}), the
characteristics of particle production are to be determined from
the conservation laws. The issue of the existence of particle
field(s) which satisfy these conditions is, however, a separate
one.

Now, consider various contributions to the energy density.
Different these contributions scale quite differently with the
temporal functions. In particular, it can be seen, following an
analysis similar to that for the FLRW spacetime, that if
$\rho_{{\rm NRM}}$ is currently dominating then $\rho_{{\rm RAD}}$
dominated in the past, while $\rho_{{\rm VAC}}$ dominated very
early during the temporal history of (\ref{genhsp}).

Thus, as is the case with the FLRW spacetime, we will obtain the
expanding ``cosmic fireball" of the Big Bang conception with the
expansion of the spacetime being driven by the decaying vacuum
energy density.

The origin of the observed \cite{partridge, cobe} Microwave
Background Radiation (MBR) is then the hot and dense ``cosmic
fireball" phase in the early history of the universe in
(\ref{genhsp}). The Big Bang Nucleosynthesis for (\ref{genhsp})
may proceed in a manner similar to that for the FLRW spacetime.

Earlier, the author was prompted to consider \cite{gen-cos} an
alternative general relativistic cosmology based on
(\ref{genhsp}).

However, the conclusive evidence \cite{srianand, subir, cmbtempz}
that the temperature of MBR was higher in the past rules out that
cosmology. It is not possible to simultaneously explain the
increase of MBR temperature out to red-shifts $z\,\sim \,3$ and
other observations when we assume that the entire red-shift is
``intrinsic" to the object and that the spacetime is
non-expanding.

We have provided here only the basic outline of the Big Bang
Cosmology of (\ref{genhsp}). Above all, however, the complexity of
the real observable universe is enormous and the details of the
present cosmology are being worked out.

Then, it is the present model, the inhomogeneous big bang model,
that appears to be satisfactory since it has no ``known" problems
of principle associated with it.

Now, since the entire matter in the present cosmology is expected
to attain the speed of light in the asymptotic future. This
cosmology, therefore, predicts ``hot death" for the universe.

\acknowledgements{ Some calculations have been performed using the
software {\tt SHEEP} and I am indebted to Malcolm MacCallum for
providing this useful software to me. I am also grateful to
Ravindra Saraykar for many discussions and, to Subir Sarkar, Bruce
Bassett and Joakim Edsjo for drawing my attention to recent
observational results on the higher MBR temperature at large
red-shifts. }

\end{document}